\documentclass[notitlepage,prx,longbibliography,letter,reprint,nofootinbib]{revtex4-1}

\usepackage{graphicx}
\usepackage{amsmath}
\usepackage{amssymb}
\usepackage{hyperref}
\usepackage{tikz}

\newcommand{\be}{\begin{eqnarray*}}
\newcommand{\ee}{\end{eqnarray*}}
\newcommand{\beq}{\begin{eqnarray}}
\newcommand{\eeq}{\end{eqnarray}}

\newcommand{\dd}{\mathrm{d}}

\newcommand{\ket}[1]{\left|{#1}\right\rangle}

\newcommand{\hc}{\mathrm{H.c.}}

\newcommand{\zbb}{\mathbb{Z}}
\newcommand{\tbb}{\mathbb{T}}
\newcommand{\id}{\mathbb{I}}

\begin{document}
\title{Abelian Floquet symmetry-protected topological phases in one dimension}
\author{Rahul Roy}
\author{Fenner Harper}
\affiliation{Department of Physics and Astronomy, University of California, Los Angeles, California 90095, USA}
\date{\today}
\begin{abstract}
Time-dependent systems have recently been shown to support novel types of topological order that cannot be realized in static systems. In this paper, we consider a range of time-dependent, interacting systems in one dimension that are protected by an Abelian symmetry group. We classify the distinct topological phases that can exist in this setting and find that they may be described by a bulk invariant associated with the unitary evolution of the closed system. In the open system, nontrivial phases correspond to the appearance of edge modes, which have signatures in the many-body quasienergy spectrum and which relate to the bulk invariant through a form of bulk-edge correspondence. We introduce simple models which realize nontrivial dynamical phases in a number of cases, and outline a loop construction that can be used to generate such phases more generally.
\end{abstract}
\maketitle
\section{Introduction}
In recent years, topological order has become a concept of fundamental importance in condensed matter physics. The most unifying work in this field has been the production of detailed classification schemes that systematically categorize the variety of topological orders that may exist in physical systems under certain conditions. The original classification schemes considered gapped, noninteracting fermionic systems, leading to the so-called ``tenfold way'' of topological insulators and superconductors \cite{Kitaev:2009vc,Ryu:2010ko}. More recently, classification schemes have been developed for interacting systems that are protected by a global symmetry, $G$, but which have no intrinsic (long-range) topological order, known as symmetry protected topological phases (SPTs) \cite{Wang:2015fi,Cheng:2015tx,Gu:2014cu,Kapustin:2014vo,Chen:2012je,Wang:2014wv,Vishwanath:2013fv,Chen:2013fo,Levin:2012jq,Pollmann:2010iha,Chen:2011gt,Fidkowski:2011dh,Schuch:2011ip,Chen:2011kg,Chen:2010gb,Fidkowski:2010ko,Pollmann:2010ih,Gu:2009ki}. Using group cohomology and other techniques, schemes have been developed that classify SPT phases for systems with many different symmetry groups and in various dimensions.

Topological or SPT order was originally thought only to exist in a strict sense at zero temperature, since many topological phases break down in the presence of thermal excitations \cite{Nussinov:2008jf,Alicki:2009hm}. Recently, however, this belief has been overturned by the theoretical prediction of many-body localisation (MBL), a phenomenon wherein strong disorder induces non-thermal behavior in a many-body system, which may in turn exhibit ordered phases that correspond to high temperatures \cite{Basko:2006hh,Oganesyan:2007ex,Pal:2010gr,Bardarson:2012gc,Serbyn:2013cl,Serbyn:2013he,Bauer:2013jw,Magan:2014sca,Huse:2014co,Nandkishore:2015kt}. In systems undergoing MBL, topological order can persist above zero temperature \cite{Huse:2013bw,Bauer:2013jw}. Moreover, in some MBL SPT systems with sufficiently strong disorder, the entire spectrum can display signatures of topological order, extending the notion of classification beyond the study of the ground state \cite{Chandran:2014dk,Bahri:2015ib,Potter:2015vn,Slagle:2015uo}.

SPT order has also been studied in the context of Floquet systems: systems described by a time-dependent Hamiltonian $H(t)$ that varies periodically in time with period $T$. In a Floquet system, to be described more fully below, the relevant states are the eigenstates of the unitary evolution operator at the end of the driving cycle, $U(T)$ \cite{Cayssol:2013gk}. Using periodic driving, it is possible to force a system into a topological phase (see Ref.~\cite{Bukov:2015gu} for a review), and in particular, to generate stroboscopic SPT phases \cite{Iadecola:2015dg}. Remarkably, Floquet systems have also been shown to exhibit novel types of topological order that cannot exist in static systems, including the existence of edge modes when the bulk Hamiltonian is trivial \cite{Kitagawa:2010bu,Lindner:2011ip,Jiang:2011cw,Thakurathi:2013dt,Rudner:2013bg,Asboth:2014bg,Carpentier:2015dn,Nathan:2015bi,Fruchart:2016hk,Roy:2016vi}. More recently, topological order in Floquet systems has been shown to persist in the presence of disorder \cite{Titum:2015fl,Titum:2015wj}.

The study of interacting phases, however, has proved to be a more difficult endeavour. Driven, interacting systems generically lead to heating, and the resulting eigenstates might naively be expected to fall into trivial, infinite-temperature, ergodic phases \cite{Lazarides:2014ie}. This may be prevented by introducing strong disorder to the system, which leads to localized phases that avoid the problems of infinite heating \cite{DAlessio:2013fv,Ponte:2015hm,Lazarides:2015jd,Ponte:2015dc,Abanin:2014te,Abanin:2015bc,Khemani:2015wn}. Moreover, as demonstrated in Ref.~\cite{Khemani:2015wn} in a model of one-dimensional (1D) spin chains with Ising symmetry, MBL allows sharp phases to be defined by considering order in the Floquet eigenstates. Two of the phases identified in this work have no static analog, and are in fact connected to Floquet topological insulators.

In this paper, we attempt to classify interacting, symmetry-protected Floquet systems, focusing mainly on Abelian symmetry groups and systems of bosons in one dimension, and building on intuition gained from studying the noninteracting problem \cite{Roy:2016vi} and the disordered Majorana Floquet chains in Ref.~\cite{Thakurathi:2013dt,Khemani:2015wn}. Our approach considers the decomposition of a general unitary evolution into a constant Hamiltonian component, which may generate topological order analogous to a static SPT phase, and a loop component, which can introduce dynamical SPT order that is only possible in driven systems. We present expressions for topological invariants and illustrate the bulk-edge correspondence for these phases, furnishing our results with examples from class D and $\zbb_N\times\zbb_N$ SPTs. Several recent papers consider the classification of Floquet SPT phases using different methods \cite{vonKeyserlingk:2016vq,Else:2016tj,Potter:2016tb}, and we find that our results are consistent with these.

The structure of this paper is as follows. We begin in Sec.~\ref{sec:PreDis} by discussing some general features of time-dependent systems and introducing the concept of a unitary loop evolution, which we will require in the main text. In Sec.~\ref{sec:ClassD} we consider fermionic systems in class D, which act as a bridge between free and interacting 
Floquet systems. We then study bosonic systems with $\zbb_N \times \zbb_N$ symmetry in Sec.~\ref{sec:ZNZN} and more general Abelian groups in Sec.~\ref{sec:GenAbe}, where we also discuss bulk invariants and bulk-edge connections. We end with a discussion in Sec.~\ref{sec:Discussion}.

\section{Preliminary Discussion\label{sec:PreDis}}
The evolution due to a local, time-dependent Hamiltonian $H(t)$ can be described by the unitary operator
\beq
U(t)&=&\mathcal{T}\exp\left(-i\int^t_0 H(t')\dd t'\right),\label{eq:unitary_operator}
\eeq
where $\mathcal{T}$ is the time ordering operator and $t$ usually runs from $0$ to $T$. We assume that the Hamiltonian $H(t)$ at every instant is invariant under a group of symmetry transformations, $G$. We write the instantaneous eigenstates of $U(t)$ as $\ket{\psi(t)}$ and note that these correspond to instantaneous quasienergies $\epsilon(t)$ through $U(t)\ket{\psi(t)}=e^{-i\epsilon(t)}\ket{\psi(t)}$, where $-\pi<\epsilon(t)\leq \pi$.  It is conventional to define the effective Floquet Hamiltonian at the end of the evolution as $H_F$, obtained through $U(T)=\exp\left(-iH_FT\right)$, when this gives a meaningful (i.e., local and symmetry-preserving) Hamiltonian. 

For the purposes of this paper, we restrict our discussion to unitary evolutions which, for a closed system, can be expressed as $\exp(-iH_FT)$ for some Floquet Hamiltonian $H_F$. As we will see below, this implies that the unitary for the open system can be written as a product of two factors, where one factor is of the form $\exp(-iH_F^oT)$ and the other is an effective edge unitary operator. We want to understand and classify the distinct types of robust effective edge operator that can arise from such unitary evolutions. The requirement that the effective edge unitary have observable physical consequences places some constraints on the form of $H_F$, which we do not study in detail here. One expects that integrability or MBL is necessary, although likely not sufficient.

One way to generate edge modes protected from bulk transitions by a quasienergy gap or an effective mobility gap in an MBL phase, is by evolving a constant SPT Hamiltonian in time. Phases produced in this way have protected edge modes, resulting in a quasienergy spectrum with degenerate multiplets at every level. The edge modes, which are connected to this vector space, then correspond to a projective representation of the group, and the different phases correspond to equivalence classes of projective representations \cite{Chen:2013fo}.

However, free fermion studies point to a set of more interesting possibilities, which we will find correspond to nontrivial \emph{unitary loops}. We define a loop to be a unitary evolution of the form in Eq.~\eqref{eq:unitary_operator} that, for a closed system, satisfies $U(T)=e^{i\phi}\id$ for some phase $\phi$. In much of what follows, we will assume $\phi=0$ without loss of generality. In this way, the quasienergies at the end of a unitary loop coincide, and the effective Floquet Hamiltonian $H_F=0$. Nevertheless, after the loop evolution of an open system, there may be edge modes present at different quasienergy values, which leads to a richer set of possible topological phases. 

The utility of studying loops becomes evident when we consider an end point unitary for a closed system that can be expressed as $U(T)=\exp(- i H_FT)$, where $H_F$ is a static and local Hamiltonian. If this is the case, then we can continuously deform the unitary evolution into a composition of unitaries of the form $U=C*L$, where $L$ is a loop and $C$ is the evolution due to the static Hamiltonian $H_F$. We define the composition of two unitaries $U_1*U_2$ in the usual way, as the complete evolution due to $U_1$ followed by the complete evolution due to $U_2$, with both components appropriately rescaled so that the evolution runs from $t=0$ to $t=T$ (see Ref.~\onlinecite{Roy:2016vi} for an explicit construction).

If the closed system unitary $U(t)$ results from the evolution due to the time-dependent Hamiltonian $H(t)$, and has the end-point form $U(T)=\exp(- i H_FT)$, the continuous deformation into the decomposed form $U=C*L$ may be obtained through the homotopy Hamiltonian
\begin{equation}\label{eq:loop_decomposition}
H'(t,s)=\left\{\begin{array}{cc}
+2H_Fs & 0 \leq t \leq Ts/2\\
-4H_Fs & Ts/2 \leq t \leq 3sT/4\\
H\left(\frac{t-3sT/4)}{1-3s/4}\right) & 3sT/4\leq t \leq T.
\end{array}\right.
\end{equation}
When $s=0$, this Hamiltonian generates the original $U(t)$, but when $s=1$, the first half of the evolution is a constant (scaled) evolution corresponding to $H_F$, and the second half of the evolution corresponds to a loop $L(t)$.\footnote{We assume in this article that the constant evolution precedes the loop evolution in the decomposition of a general unitary. Similar arguments may be obtained using the opposite ordering.} The end point of the evolution remains fixed throughout. In a noninteracting system, this unitary decomposition leads to a rigorous classification of unitary evolutions \cite{Roy:2016vi}.

In an open system, the unitary evolution will differ from that of the closed system at the edge, giving rise to an effective edge unitary. Notably, for a nontrivial state, it may not be possible to write the endpoint unitary for the open system as $U^o(T)=\exp\left(-iH_F^oT\right)$ for a local, symmetry-preserving Hamiltonian $H_F^o$: some terms could connect sites at the two edges of the chain. In this case, following the decomposition of Eq.~\eqref{eq:loop_decomposition}, the open system unitary takes the form $U^o(T)=U_{\rm eff}\exp\left(-iH_F^oT\right)$, where $U_{\rm eff}$ is the effective edge unitary that derives from the loop component of the evolution.

Edge unitaries of loops have some special properties, as we shall see below, and these are likely to persist when the loops are followed by unitaries with local integrals of motion. When the constant Hamiltonian is MBL, the SPT classification of 1D ground states \cite{Chen:2010gb,Fidkowski:2010ko,Chen:2011kg,Pollmann:2010iha} can be extended to all eigenstates \cite{Huse:2013bw,Chandran:2014dk} with detectable consequences \cite{Chandran:2014dk,Bahri:2015ib}.\footnote{We note, however, that robust edges may also exist for systems without MBL \cite{Fendley:2015ie}.} The effective edge unitary of a loop followed by a constant MBL Hamiltonian of some SPT phase will not in general have the direct product structure of bulk times edge. However, it is likely that the edge modes of the loop will continue to have some observable signatures, such as those considered in Ref.~\onlinecite{Bahri:2015ib}. The classification of possible phases of such systems, as obtained from studying the distinct edge unitaries, then has a natural product structure, $n(L) \times H^2(G,U(1))$ where the second factor originates from the usual static classification of SPT phases~\cite{Chen:2012je,Chen:2013fo} and the first factor counts the number of distinct loops. 

Since the constant evolution component is well understood, we will focus our study on loops. As noticed previously, generic Floquet systems driven with interacting Hamiltonians suffer from heating, and this is certainly true for loops, which are featureless and hence, already at infinite temperature. However, one can construct an MBL Floquet drive which avoids this problem by pairing a loop with a constant MBL Hamiltonian. More generally, loops are of interest and may be defined even in systems without MBL. In this work, we restrict ourselves to one dimension where progress can be made most easily. We study nontrivial loops in a variety of contexts and characterize them by bulk topological invariants.

\section{Class D Systems\label{sec:ClassD}}
To motivate the ideas described in this paper, we begin by studying the example of class D free fermionic systems \cite{Thakurathi:2013dt,Jiang:2011cw}. These have a simple description in terms of noninteracting Majorana fermions, and allow us to develop the intuition required for the many-body picture that we will use in later sections.

\subsection{Single-particle Picture\label{sec:ClassD_singlep}}
As motivated in Sec.~\ref{sec:PreDis}, we will focus on forming nontrivial unitary loops, which will in turn allow us to generate systems with dynamical topological order that are unique to time-dependent systems. In a free-fermion system, loops of this kind can always be generated by evolving with a nontrivial, flat-band SPT Hamiltonian for the first half of a cycle, and a trivial, flat-band Hamiltonian for the remainder of the cycle. We will illustrate this construction with a specific example below.

A general class D Hamiltonian may be written in terms of Majorana fermions \cite{Kitaev:2001gb} as
\beq
H&=&\frac{i}{4}\sum_{ij}\gamma_iA_{ij}\gamma_j,\label{eq:maj_second_quant_ham}
\eeq
where $\gamma_i^\dagger=\gamma_i$ are Majorana fermions and $A_{ij}$ is an antisymmetric matrix. In the static system, this Hamiltonian is classified by studying the symmetric momentum points, $k=0$ and $k=\pi$, about which the Hamiltonian has the full symmetry of class D. By considering the appropriate homotopy groups, it can be shown that static systems in class D have a twofold classification described by an element of $\zbb_2$ \cite{Kitaev:2001gb}.

In the Floquet case, we form (and define) the Majorana unitary evolution operator through $O(t)=\mathcal{T}\exp\left(\int_0^tA_{ij}(t')\dd t'\right)$. Since $A_{ij}$ is antisymmetric, and $O(0)$ is the identity, we see that $O(t)$ belongs to the special orthogonal group $SO(2n)$, if $2n$ is the number of bands in the BdG Hamiltonian. 

We can classify the Floquet loop by again studying the behavior at the symmetric momentum points, $k=0$ and $k=\pi$. Since we are considering loops, $O(t)$ must end at the identity matrix, and so the classification is given by the fundamental group $\pi_1\left(SO(2n)\right)$, which for $n>1$ is $\zbb_2$. We see that there is a $\zbb_2$ invariant associated with both $k=0$ and $k=\pi$, yielding four possible phases in total. The set of loops is therefore classified by $(e_1,e_2)\in\mathbb{Z}_2\times\mathbb{Z}_2$ for $n>1$, with each $e_i$ taking values from $\{0,1\}$. 

The topological invariant that determines the existence (or lack) of nontrivial boundary modes for an open system is given by the sum (modulo 2) $e_1+e_2\in\mathbb{Z}_2$. The extra $\mathbb{Z}_2$ element, which does not correspond to the edge modes, may be regarded as a weak topological invariant of the drive.\footnote{Note that for $n=1$, the classification of loops is $\mathbb{Z}\times\mathbb{Z}$, but the $\mathbb{Z}_2$ invariant that describes the edge is still given by $\left(e_1+ e_2\right)\mod{2}$.} The nontrivial Floquet loop is characterized by a single Majorana $\pi$ mode in the open system, which, in the absence of additional static SPT order, is also accompanied by a Majorana zero mode. For a unitary evolution that is not a loop, there may also be edge mode contributions from a subsequent constant Hamiltonian evolution, as described in Sec.~\ref{sec:PreDis}.

We now illustrate these ideas by describing a specific model in class D, with further details given in Appendix~\ref{app:classD_details}. We consider a one-dimensional fermionic chain of length $K$ that has a two-state Hilbert space at each site, with the corresponding annihilation operators on site $j$ being $a_j$ and $b_j$.\footnote{We choose a two-state Hilbert space so that the resulting phase has nontrivial dynamical SPT order but trivial static SPT order. This requires a Majorana mode at $\epsilon=0$ and a Majorana mode at $\epsilon=\pi$.} Following Kitaev \cite{Kitaev:2001gb}, we can define two sets of Majorana operators through
\beq
\gamma^a_{2j-1}=a_j+a^\dagger_j,&~~~~~~&\gamma^a_{2j}=\frac{a_j-a^\dagger_j}{i}\nonumber\\
\gamma^b_{2j-1}=b_j+b^\dagger_j,&~~~~~~&\gamma^b_{2j}=\frac{b_j-b^\dagger_j}{i},
\eeq
which satisfy $\gamma^\dagger=\gamma$. Let
\beq
H_{1a}&=&-\sum_j\left(a^\dagger_ja_j-\frac{1}{2}\right)=-\frac{i}{2}\sum_j \left(\gamma^a_{2j-1}\gamma^a_{2j}\right)\nonumber\\
H_{2a}&=&\frac{1}{2}\sum_j\left(-a^\dagger_ja_{j+1}-a^\dagger_{j+1}a_j+a_{j}a_{j+1}+a_{j+1}^\dagger a^\dagger_j\right)\nonumber\\
&=&\frac{i}{2}\sum_{j}\left(\gamma^a_{2j}\gamma^a_{2j+1}\right),\label{eq:classD_Ham}
\eeq
with $H_{1b},H_{2b}$ defined identically in terms of the $b$ operators and $\gamma^b$ Majoranas. $H_{1a}$ and $H_{2a}$ respectively correspond to the trivial and nontrivial phases of the 1D class D superconductor.

We now evolve the system with $H_1 = H_{1a}+2H_{1b}$ for $0\leq t\leq \pi$ and with $H_2=H_{2a}+2H_{2b}$ for $\pi\leq t\leq 2\pi$. We notice that for an open system, $H_{2a}$ and $H_{2b}$ each have Majorana edge modes. The evolution by the trivial Hamiltonian $H_{1a}$ pushes the Majorana mode of subsystem $a$ to quasienergy $\epsilon=\pi$, while the Majorana mode of subsystem $b$ has been pushed to quasienergy $\epsilon =2\pi\equiv 0$.  Thus, we see that at $t=\pi$, a Majorana mode at both zero and $\pi$ emerge, which persist until the end of the evolution. On the other hand, evolving the closed system until $t=2\pi$ leads to a unitary that is the identity (up to an overall phase factor). In this way, the two-part drive described by Hamiltonians $H_1$ and $H_2$ produces a nontrivial unitary loop with dynamical SPT order. 

This order is manifested in the properties of the unitary evolution operator. At the end of the loop, the unitary describing the closed system differs from the unitary describing the open system through terms that act at the ends of the chain. In the Majorana language, it is clear that the term $\frac{i}{2}\gamma^a_{2K}\gamma^a_1$ from $H_{2a}$ and the term $\frac{i}{2}\gamma^b_{2K}\gamma^b_1$ from $H_{2b}$ cannot act in the open system, since they would connect the two edges. If we define new fermion operators through
\beq
d=\frac{1}{2}\left(\gamma^a_{2K}+i\gamma^a_1\right),&~~~~~~&d^\dagger=\frac{1}{2}\left(\gamma^a_{2K}-i\gamma^a_1\right)\nonumber\\
e=\frac{1}{2}\left(\gamma^b_{2K}+i\gamma^b_1\right),&~~~~~~&e^\dagger=\frac{1}{2}\left(\gamma^b_{2K}-i\gamma^b_1\right),
\eeq
we see that the open and closed unitary operators ($U_{\rm op}$ and $U_{\rm cl}$, respectively) are related through
\beq
U_{\rm op}(2\pi)&=&e^{\left[-\frac{\pi}{2}\left(\gamma^a_{2K}\gamma^a_1+2\gamma^b_{2K}\gamma^b_1\right)\right]}U_{\rm cl}(2\pi)\\
&=&e^{\left[\frac{i \pi}{2}\left(d^\dagger d- dd^\dagger\right)+{i \pi}\left(e^\dagger e- ee^\dagger\right)\right]}U_{\rm cl}(2\pi),\nonumber
\eeq
where $U_{\rm cl}(2\pi)$ is the identity up to some overall phase factor. The details of this calculation are presented in Appendix~\ref{app:classD_details}. Thus, we see that the effective unitary at the edge is 
\beq
U_{\rm eff}(2\pi)&=&e^{\left[\frac{i \pi}{2}\left(d^\dagger d- dd^\dagger\right)+{i \pi}\left(e^\dagger e- ee^\dagger\right)\right]},\label{eq:classD_eff_unit}
\eeq
where the second term in the exponent merely provides an overall factor of $-1$.

We will use the concept of an effective edge unitary throughout this article to discuss topological order and define invariants. We will be particularly interested in its behavior under the symmetry transformations of the relevant group $G$, which for class D is fermion parity. We denote the parity operator by $\hat{P}$, which can be expressed as a product of the onsite parity operators of all sites, $\hat{P}=\prod_{j}\hat{P}_j$, with $\hat{P}_j=\left(-i\gamma_{2j-1}^a\gamma_{2j}^a\right)\left(-i\gamma_{2j-1}^b\gamma_{2j}^b\right)$. Now, if we define an effective parity operator $\hat{P}_{L}$ at the left edge of the open chain, which is a product over a finite set of onsite parities in the vicinity of this edge, then for the effective edge unitary in Eq.~\eqref{eq:classD_eff_unit}, we find
\beq
\{\hat{P}_L , U_{\rm eff}\}&=& 0. \label{eq:classd_eff_unit_char}
\eeq
For a trivial effective edge unitary, we would instead find  $[\hat{P}_L , U_{\rm eff}]=0 $.

In the discussion above, we have considered a specific model Hamiltonian, whose unitary evolution exhibits clear signatures of the underlying dynamical SPT order. While this model was chosen for its simplicity, the resulting signatures will apply more generally. First, a Majorana $\pi$ mode will exist at any point after a dynamical phase transition has occurred (until another equivalent phase transition occurs). In the model considered here, the existence of the $\pi$ mode could be shown analytically at the special point $t=\pi$. At other times, and for more general models, the presence of a Majorana $\pi$ mode may be inferred from the properties of the many-body spectrum, which we discuss below in Sec.~\ref{sec:phase_transitions}.

We may also consider perturbing the system with a local (short-range) Hamiltonian, $H_L$, which acts near the left edge at the end of the loop (and which hence commutes with a suitably defined $\hat{P}_L$). After time $t$, the effective edge Hamiltonian will have evolved to $U_{\rm eff}'=e^{-iH_Lt}U_{\rm eff}$. However, it is clear that the new effective unitary still satisfies $\{\hat{P}_L , U_{\rm eff}' \}=0$. In this way, the effective edge unitary is characterized by its commutation or anticommutation properties with the parity operator, $\hat{P}_L$, and is robust under local perturbations that preserve symmetry.

We can also consider threading a $\pi$-flux through the closed system, and the effects of this will provide another signature of the SPT order which is robust to perturbations. By explicit calculation, detailed in Appendix~\ref{app:classD_details}, it can be shown that the unitary of the closed system at the end of the cycle with flux differs from the unitary without flux by an overall factor of $e^{i\pi}$. We write this relationship as $U_0U^\dagger_\pi=e^{i\pi} \mathbb{I}$. 
\subsection{Many-body Picture}
We are now in a position to formulate the problem in a many-body setting, returning to the general set of systems in class D. Although we will at first not introduce any additional interactions, we note that much of the discussion that follows is applicable to both free and interacting systems. This is in much the same way that the classification of static class D free-fermion systems carries over to the interacting case \cite{Kitaev:2001gb}.
\subsubsection{Bulk Invariant}
To find a bulk invariant in the many-body picture, we consider the unitary evolution corresponding to the general Hamiltonian given in Eq.~\eqref{eq:maj_second_quant_ham}. A nontrivial Majorana Floquet loop satisfying $O(T)=\mathbb{I}$ corresponds to a fermionic unitary loop satisfying $U(T)=-\mathbb{I}$. For a nontrivial evolution, we therefore see that $U_0(T)U^\dagger_\pi(T)=e^{i\pi}\mathbb{I}$, which agrees with our findings for the specific model considered in the previous section. For a trivial loop, the equivalent product yields $U_0(T)U^\dagger _\pi(T)=\mathbb{I}$, and so we can define the bulk invariant, $k \in \{0,1\} \equiv \mathbb{Z}_2$ of an interacting class D system through $U_0(T)U^\dagger _\pi(T) = e^{ik\pi} \mathbb{I}$. We expect this invariant to characterize a loop even in the presence of parity-preserving interactions, and we will argue that this is indeed the case below.

\subsubsection{Edge Picture}
We found previously that a nontrivial unitary loop corresponds to the appearance of a zero and a $\pi$ Majorana mode in the single-particle spectrum of an open system. In the many-body spectrum, this is manifested as a twofold degeneracy at every quasienergy, accompanied by a $\pi$-degeneracy. By $\pi$-degeneracy, we mean that every state in the spectrum has a counterpart separated by a quasienergy $\pi$, which differs only by a boundary Majorana mode. This implies that there are two states of differing parity at each energy, which we expect to persist in the presence of parity-preserving interactions. 

Without loss of generality, we will restrict our discussion to nontrivial loops that can be written as a two-part drive, comprising evolution by a trivial Hamiltonian followed by evolution by an SPT Hamiltonian (more general unitary loops can be considered as compositions of such two-part drives). In these cases, the unitary for the closed system is the identity (up to a phase), while the nontrivial behavior at the edge of the open system can be captured by an effective unitary of the form given in Eq.~\eqref{eq:classD_eff_unit},
\beq
U_{\rm eff}&=&e^{\left[\frac{i\pi}{2}\left(d^\dagger d -dd^\dagger \right)+i\pi\left(e^\dagger e -ee^\dagger \right)\right]}.\label{eq:classD_mb_eff_unit}
\eeq
In the general interacting case, the operators $d$ and $e$ can be many-body fermion operators that depend nontrivially on the operators of the original Hamiltonian, but which transform under parity in the same way as in the single-particle case. We regard the $\pi$- and 0-Majorana modes as arising out of a dynamical phase transition that does not change the static SPT order of the system.

\subsubsection{Phase Transitions\label{sec:phase_transitions}}
We now consider the phase transitions that must accompany changes in the dynamical topological order of the system. Formally, we note that a single unitary evolution $U(t)$ for $t\leq T$ can also be regarded as a family of unitaries $U'(t,s)$ with
\be
U'(t,s) &=& U(st)
\ee
for $st\leq T$ and where $s$ is continuously varied. 

If, as in the model system above, the closed system unitary $U'(T,s)$ can be expressed as the exponent of a natural Floquet Hamiltonian for values of $s$ excluding some point $s_c$, but the open system unitary does not have such a natural Floquet Hamiltonian expression for $s>s_c$, then we may regard this as a phase transition of the unitary that occurs at $s_c$. The transition point $U'(T,s_c)$ is then equivalent to $U(T/s_c)$. We will loosely refer to this as the existence of a critical point in the original Floquet evolution $U(t)$ at the point $t=T/s_c\equiv t_c$. In other words, we assume that for $t>t_c$ the open system unitary takes the form $U^o(t>t_c)=U_{\rm eff}\exp(-iH_F^ot)$, with the dynamical order characterized by the effective edge unitary $U_{\rm eff}$.

We assume that the system undergoes such a transition at $t=t_c$, and consider the instantaneous many-body quasienergy spectrum of $U(t)$ across $t_c$.  A similar approach was used to study Floquet transitions in the noninteracting case in Ref.~\cite{Nathan:2015bi}. We assume that the static SPT order of the eigenstates of the unitary evolution operators before and after the transition is trivial, so that there are no Majorana modes before $t_c$ and one Majorana mode each at zero and $\pi$ after $t_c$. For simplicity, we will also assume that there are no other phase transitions during the evolution. These assumptions suggest that there is no difference between the closed and open systems for $t<t_c$, since in this region the evolution is topologically trivial.

At the transition point $t_c$, the emergence of Majoranas is associated with a bulk $(0,\pi)$-transition, where each bulk eigenstate $\ket{\psi}$ becomes part of a multiplet of four states that may be written $\{\ket{\psi},d^\dagger\ket{\psi},e^\dagger\ket{\psi},d^\dagger e^\dagger\ket{\psi}\}$ after the transition. If two of these states have many-body quasienergy $\epsilon (t_c)$,  the other two have quasienergy $\epsilon (t_c)+\pi$ (modulo $2\pi$). The two states in each pair have opposite parity. This degeneracy pattern in the many-body spectrum will persist until another phase transition occurs.

\begin{figure}
\includegraphics[scale=0.48]{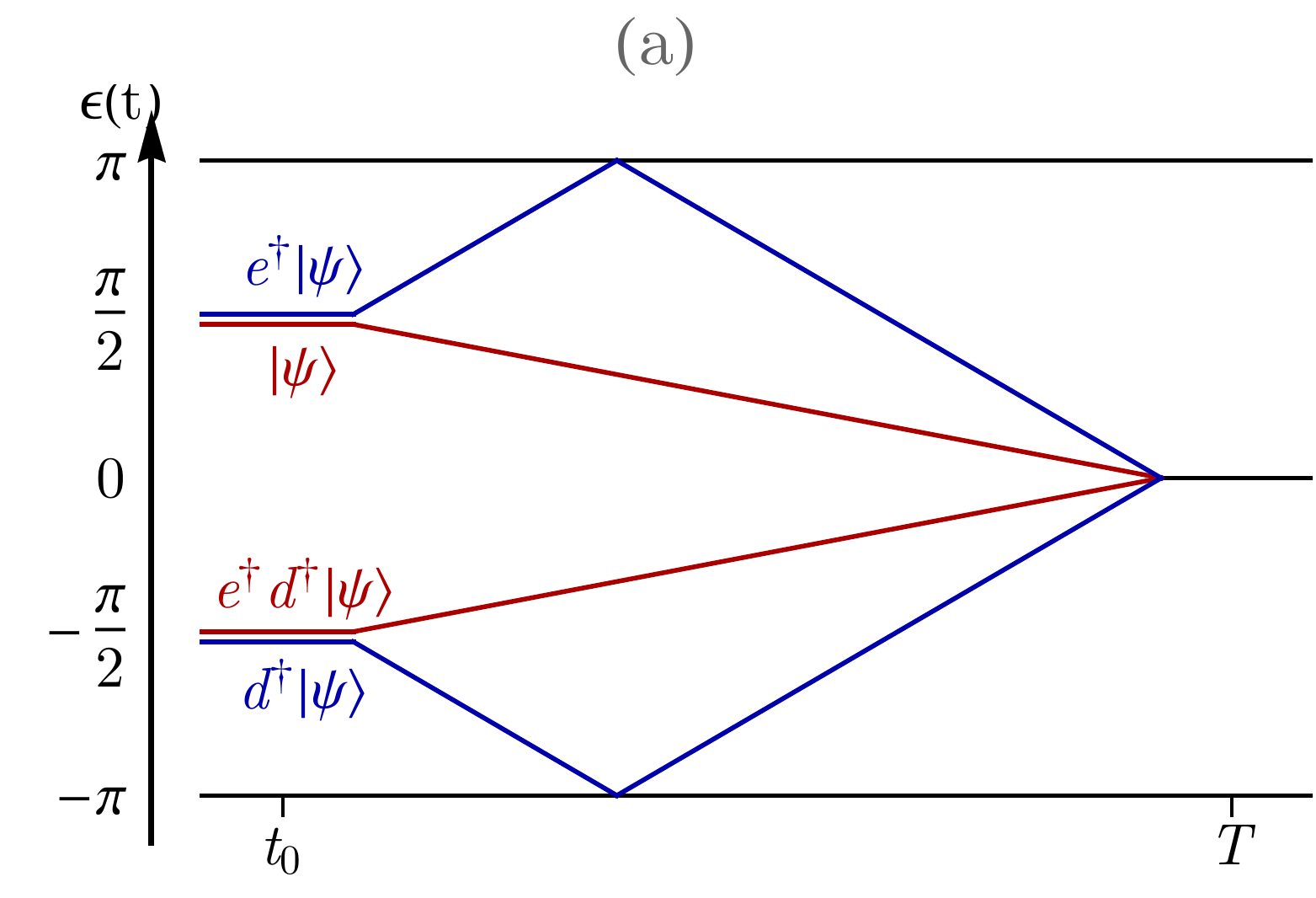}
\includegraphics[scale=0.48]{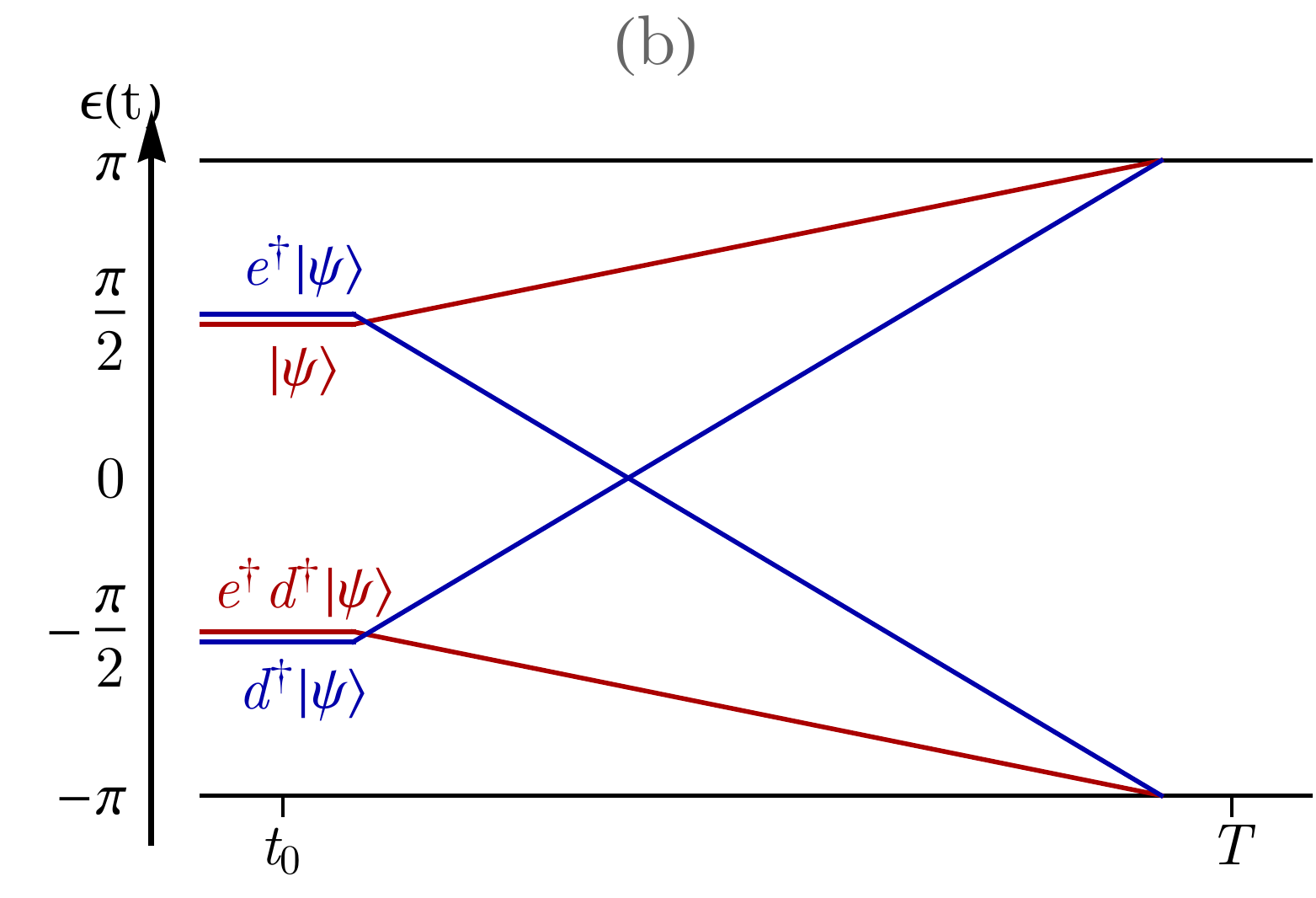}
\caption{Schematic splitting of a Majorana multiplet at a $(0,\pi)$-phase transition as the system is closed with (a) periodic and (b) antiperiodic boundary conditions. Blue lines correspond to states with even parity, while red lines correspond to states with odd parity. Spectrum (b) can be obtained from spectrum (a) through a shift by $\pi$. See main text for details.\label{fig:ClassDMultiplet}}
\end{figure}

We now consider what happens in the open system if we reconnect the edges at some point just beyond $t_c$ with a Hamiltonian of the form
\beq
H'&=&h_d\left(d^\dagger d-dd^\dagger \right)+h_e\left(e^\dagger e-ee^\dagger \right).\label{eq:edge_closing_ham}
\eeq
The effective edge unitary at time $t>t_c$ is then given by
\beq
U_{\rm eff}'(t)&=&e^{\left[\frac{i\pi}{2}\left(d^\dagger d -dd^\dagger \right)-i(t-t_c)h_d \left(d^\dagger d-dd^\dagger \right)\right]}\times\nonumber\\
&&e^{\left[i\pi\left(e^\dagger e-ee^\dagger \right)-i(t-t_c)h_e \left( e^\dagger e-ee^\dagger \right)\right]},
\eeq
which is the effective unitary from Eq.~\eqref{eq:classD_mb_eff_unit} multiplied by the evolution due to the edge-closing Hamiltonian, Eq.~\eqref{eq:edge_closing_ham}. We note that if we close the system with antiperiodic boundary conditions, the sign of the coefficients $h_d$ and $h_e$ changes relative to closing the system with periodic boundary conditions. This allows us to make the following statements about the spectrum immediately after $t_c$:
\begin{enumerate}
\item	In the system with a $\pi$ flux (which is equivalent to imposing antiperiodic boundary conditions), the spectrum after $t_c$ may be obtained from the system with zero flux by shifting the entire spectrum by $\pi$.
\item	If the zero flux spectrum is shifted in this way, then states are mapped onto states with the same parity.
\item	If $h_e\neq0$, there is a parity shift in the ground state (or any other particular state) of the multiplet in the system with flux.
\end{enumerate}
These spectral features will arise in a generic class D system with dynamical SPT order. In Fig.~\ref{fig:ClassDMultiplet}, we show the splitting of the multiplet schematically for both periodic and anti-periodic boundary conditions, assuming the complete evolution forms a unitary loop. We note that if there are no other phase transitions, then $U_\pi(T)U^\dagger_0(T)=-\mathbb{I}$, demonstrating the correspondence between the bulk invariant and the behavior of the edge modes. 

Finally, we note that a $(0,\pi)$-transition of this form can be viewed as a fourfold energy level crossing in the folded quasienergy spectrum, which plots many-body quasienergy values modulo $\pi$ in the range $0\leq \epsilon (t) \leq \pi$. Notably, ordinary phase transitions would appear as pairwise crossings in both the folded and the unfolded spectra.

\section{Systems with $\mathbb{Z}_N\times\mathbb{Z}_N$ Symmetry\label{sec:ZNZN}}
Let us now consider Floquet phases of bosonic systems that are symmetric under the group $\mathbb{Z}_N\times\mathbb{Z}_N$, following a similar strategy to the discussion of class D. We will first construct appropriate flattened (in this context, integer-valued) SPT and trivial Hamiltonians, which have useful properties under time evolution. Constructing loops out of these, we will find nontrivial topological phases where there are quasienergy degeneracies at multiples of $2\pi/N$, related to edge modes in the open system. 

In the static case, the group cohomology classification predicts $N$ different SPT phases, which we label by some integer $c\in \{0,1,2,\ldots,N-1\}$. As shown in Appendix~\ref{app:ZNZN}, we can construct integer-valued SPT Hamiltonians $H^c_{a,b}$, defined on a spin chain with $K$ sites, which generate these static SPT phases and which have the following properties under time evolution: For a closed system,
\begin{equation}
U\left(\frac{2\pi}{N}\right)=e^{\left(-\frac{2\pi i}{N}H^c_{a,b}\right)}=\left[V\left(\mathbb{T}_o\right)\right]^{-a} \left[V\left(\mathbb{T}_e\right)\right]^{-b},
\end{equation}
where $\mathbb{T}_o,\mathbb{T}_e$ are generators of the two $\mathbb{Z}_N$ groups (corresponding to odd or even sites), and $V(\mathbb{T}_{o/e})$ are the global unitary operators corresponding to $\mathbb{T}_{o/e}$ (defined explicitly in Appendix~\ref{app:ZNZN}).

We can also form trivial Hamiltonians, $H^{\rm triv}_{a,b}$ which have the property (for both open and closed systems) that
\begin{equation}
U\left(\frac{2\pi}{N}\right)=e^{\left(-\frac{2\pi i}{N}H^{\rm triv}_{a,b}\right)}=\left[V\left(\mathbb{T}_o\right)\right]^{-a}\left[V\left(\mathbb{T}_e\right)\right]^{-b}.
\end{equation}
By construction, we can form a loop by evolving the system in time with the Hamiltonian
\begin{equation}
H(t)=\left\{\begin{array}{ccc}
-H^c_{a,b} &&\mbox{for $0\leq t\leq2\pi/N$},\\
-H^{\rm triv}_{\bar{a},\bar{b}} &&\mbox{for $2\pi/N\leq t\leq4\pi/N$},
\end{array}\right.
\end{equation}
where $\bar{a}=N-a$ is the inverse element of $a$. Choosing $c=1$ (we drop the superscript $c$ for the rest of this section), we find that the result of this evolution for an \emph{open} system is described by the effective edge unitary
\beq
U_{\rm eff}&=&\exp\left[-i\frac{2\pi}{N}\left(aH_1+bH_K\right)\right],
\eeq
where $H_1$ and $H_K$ are operators acting on the two ends of the open system, whose explicit form is given in Appendix~\ref{app:ZNZN}. Then, every state in the open system with quasi-energy $\epsilon$ is accompanied by states that differ by edge modes and which have quasienergies $\epsilon + 2\pi (k_{1} a/N + k_{2} b/N)$, where $k_{1/2}\in\{1,2,\ldots N-1\}$. This is a signature of the topological phase that will appear in the many-body spectrum.

 Analogous to the class D discussion, we may also consider the effect of twisting the boundary conditions of the closed system. In our model, the effect of $\mathbb{T}_o$-twisted boundary conditions can be realized by transforming the operators $H_1, H_K $ by $
H_1 \to X_1H_1X_1^\dagger, H_K \to X_1H_KX_1^\dagger $, where $X_1$ is the  unitary operator corresponding to the symmetry $\mathbb{T}_o$ at site 1. 

We now consider evolving the open system until $t=4\pi/N$, before reconnecting the ends of the chain with an operator $H'$. The effective unitary at the edge after this point is then given by
\beq
U'_{\rm eff}(t)&=&e^{\left[-i\left(t-4\pi/N\right)H'\right]}e^{\left[-\frac{2i\pi}{N}\left(aH_1+bH_K\right)\right]}.
\eeq
On the other hand, if we reconnect the system with $\mathbb{T}_o$-twisted boundary conditions, the Hamiltonian $H'$ changes to $ H' \to X_1H'X_1^\dagger $ and the resulting effective unitary changes to $U_{\rm eff}'(t)\to U_{\mathbb{T}_o}(t)$ where
\begin{equation}
U_{\mathbb{T}_o}(t)=X_1e^{\left[-i(t-4\pi/N)H'\right]}X_1^\dagger e^{\left[-\frac{2i\pi}{N}\left(aH_1+bH_K\right)\right]}.
\end{equation}
Since $\det (X_1)$ is non-vanishing, and
\beq
X_1^\dagger H_K X_1&=&H_K-\mathbb{I}+NP_0,
\eeq
where $P_0$ is the zero-energy subspace of $H_K$, we see that $U_{\mathbb{T}_o}(t)$ has the same spectrum as
\begin{equation}
\tilde{U}_{\mathbb{T}_o}(t)=e^{\left[-i(t-4\pi/N)H'\right]}e^{\left[-\frac{2i\pi}{N}\left(aH_1 +b(H_K- \mathbb{I}) \right)\right]},
\end{equation}
which is shifted by a phase $2\pi b/N$ compared to the untwisted unitary operator. Similarly, under $\mathbb{T}_e$-twisted boundary conditions, the spectrum of the effective unitary is shifted by $2\pi a/N$, and a general effective unitary may be characterized by these phases.

A related calculation tells us that evolving the closed system through the \textit{entire} loop with $\tbb_o$-twisted boundary conditions gives a unitary $e^{2i\pi b/N} \mathbb{I}$, while evolving with $\tbb_e$-twisted boundary conditions gives a unitary $e^{2i\pi a/N} \mathbb{I}$. As we will argue in the next section, these phases define a bulk topological invariant of the evolution. Since there are $N$ possible values for each phase, there are a total of $N^2$ distinct unitary loops, corresponding to $N^2$ distinct topological phases.

\section{General Abelian Groups\label{sec:GenAbe}}

For a general finite Abelian group with $p$ factors, $G=\zbb_{n_1}\times\zbb_{n_2}\times\ldots\zbb_{n_p}$, we follow the same strategy as before to construct models for loops with dynamical topological order. We first form an expanded group, $G'$, which contains $G$ as a subgroup, for which a nontrivial SPT can be constructed whose ground state multiplet contains states that can be labeled by all the irreducible representations of the group $G$. 

SPTs in one dimension are classified by the second cohomology group $H^2(G,U(1))$. Using the result \cite{karpilovsky1992group}
\beq
H^2(G,U(1))\cong&\prod_{1\leq j<k\leq p}\zbb_{(n_j,n_k)},
\eeq
where $(n_j,n_k)$ indicates the highest common factor of $n_j$ and $n_k$, we see that the group  $G'=\zbb_{n_1}\times\zbb_{n_2}\times\ldots\zbb_{n_p}\times\zbb_{\tilde{n}}$, where $\tilde{n}$ is the lowest common multiple (LCM) of the set $\{n_1,\ldots n_p\}$, has SPT states with ground state multiplets that contain all the irreducible representations of $G$. When coupled with appropriate trivial Hamiltonians to form loops, in much the same way as in the $\zbb_N\times\zbb_N$ case, these lead to systems whose effective edge unitaries for evolution on an open chain can be characterized by the phases $\{\alpha_j\}$, where each $\alpha_j$ labels an irreducible representation of $\mathbb{T}_j$, the generator corresponding to the group factor $\mathbb{Z}_{n_j}$. These models are characterized by a set of robust edge modes in the open system, which can be described by an effective edge unitary $U_{\rm eff}$. For a loop, since $H_F=0$, this implies that $U =U_{\rm eff}$.

Following our previous discussion, we introduce twisted boundary conditions as follows: first imagine cutting a closed chain at some site, and labeling the two edges $L$ and $R$. There exists a set of terms in the Hamiltonian for the closed chain that connect the two edges $L$ and $R$, whose sum we label by $H_{\rm edge}$.
For each generator $\tbb_j $, there is then an operator $X_{L,j}=\prod_{i\in L} V_i(\mathbb{T}_j)$, which is a product over the onsite unitary operators corresponding to $\mathbb{T}_j$ over a finite set of sites on the left edge, such that the effect of the twisted boundary conditions is  equivalent to transforming $H_{\rm edge}$ as $H_{\rm edge}\to X_{L,j}H_{\rm edge}\left(X_{L,j}\right)^\dagger$.

\subsection{Effective Edge Unitaries and Spectral Shift}
For the models we have just introduced, one can show, generalizing the arguments of the $\zbb_N\times\zbb_N$ case, that the effective edge unitaries satisfy the property
\beq
\left(X_{L,j}\right)^\dagger U_{\rm eff}X_{L,j}&=&e^{i\alpha_j}U_{\rm eff}.\label{eq:general_abelian_ueff}
\eeq
 We will argue below that this property is true more generally: Any effective edge unitary of a loop must satisfy an equation of the form of Eq.~\eqref{eq:general_abelian_ueff} for some $X_{L,j}$ which has support only at the left edge. 

Eq.~\eqref{eq:general_abelian_ueff} implies that if $\epsilon$ is a quasienergy of $U_{\rm eff}$ then so too is $\epsilon+n\alpha_j$, for any integer $n$. Furthermore, we see that $\alpha_j$ must in general be a multiple of $2\pi/n_j$, where $n_j$ is the size of the Abelian group associated with $\tbb_j$, for any $U_{\rm eff}$ that satisfies Eq.~\eqref{eq:general_abelian_ueff}.
We can label an effective unitary by the set of phases $\{\alpha_j\}$ it is multiplied by under transformation with the $\{X_{L,j}\}$, writing $U_{\rm eff}(\{\alpha_j\})$. We can thus obtain a distinct $U_{\rm eff}$ for every irreducible representation of the group $G$. The operator $U_{\rm eff}$ must also commute with $V(\mathbb{T}_j)=\prod_{i} V_i(\mathbb{T}_j)$ for any $j$, where the product is now over all sites.

Let us now study the implications of the effective unitary for the system. Suppose the effective edge unitary at some time is $U_{\rm eff}$. We then consider connecting the system at at this point and evolving it for an infinitesimal time $t$. The resulting evolution is captured by the transformation $U_{\rm eff}\to e^{-iH't}U_{\rm eff}$, where $H'$ is a Hamiltonian that has support on both $L$ and $R$.

The effect of carrying out the same process with $\mathbb{T}_j$-periodic boundary conditions is $U_{\rm eff}\to X_{L,j}e^{-iH't}X_{L,j}^\dagger U_{\rm eff}$. From Eq.~\eqref{eq:general_abelian_ueff}, this has the same spectrum as $e^{-iH't}U_{\rm eff}$ but is shifted by $\alpha_j$. Since the unitary of the full open system can be written as $U^o(T)=U_{\rm eff}\exp\left({-iH_F^oT}\right)$,
the spectrum of the full system under $\mathbb{T}_j$-periodic boundary conditions is shifted relative to the untwisted case by $\alpha_j$. 

Now consider the effect of perturbing the system with some local (short-ranged) term (with support only on $L$), described by Hamiltonian $H_L$. At time $t$ after the perturbation is applied, the effective unitary $U_{\rm eff}$ has evolved to $U'_{\rm eff}=\exp(-iH_Lt)U_{\rm eff}$. Since $H_L$ is symmetric under $V(\mathbb{T}_j)=\prod_{i}V_i(\mathbb{T}_j)$, then $V(\mathbb{T}_j)e^{-iH_Lt}\left[V(\mathbb{T}_j)\right]^\dagger=e^{-iH_Lt}$. However, since $H_L$ only has support on $L$, we see that 
\beq
V(\mathbb{T}_j)e^{-iH_Lt}\left[V(\mathbb{T}_j)\right]^\dagger&=&X_{L,j}e^{-iH_Lt}X_{L,j}^\dagger\nonumber\\
&=&e^{-iH_Lt}.
\eeq
It follows that
\beq
X^\dagger_{L,j} U'_{\rm eff} X_{L,j} &=&e^{i\alpha_j}U'_{\rm eff}.
\eeq
Thus we see that the defining property of the effective unitary (given in Eq.~\eqref{eq:general_abelian_ueff}) is stable against local perturbations.

\subsection{Phase Transitions}

Suppose we have a unitary evolution such that for some range of times $t$, the system can be described by effective edge unitaries $U_{\rm eff}(\{\alpha_j\},t)$ (and instantaneous bulk Floquet Hamiltonians $H_F(t)$) such that the effective edge unitary is nontrivial at some time $t_1$ and trivial at a later time $t_2$. As we now argue, a bulk phase transition must then have occurred at some point $t_c$ between $t_1$ and $t_2$.\footnote{We are considering the idealized situation of a phase transition occurring at a single point. More generally, extended states may persist for some interval of time, but our discussion can be easily extended to this more general case.}

We first consider evolving the open system until time $t_1$, before connecting the edges of the system and evolving it further for an infinitesimal time $\delta$, with both $\mathbb{T}_j$-twisted and regular boundary conditions. It follows that the spectrum of the transformed edge unitary at $t_1+\delta$ in the twisted case is shifted by $\alpha_j$ relative to the untwisted case. On the other hand, if we evolve the open system until $t_2$, and then again evolve it for an infinitesimal time $\delta$, closing with twisted and regular boundary conditions, there is no relative spectral shift between the two boundary conditions.
   
This change in the boundary behavior implies a bulk phase transition at some point $t_c$ between these two times, where by continuity, the bulk spectrum at $t_c$ is symmetric under shifts by $\alpha_j$. At the phase transition point, the edge modes at the two ends of the open system are connected through bulk modes. This transition is of the type nontrivial to trivial, but it is clear that a similar process occurs for transitions of the type trivial to nontrivial. At a phase transition where a nontrivial $U_{\rm eff}$ characterized by $\{\alpha_j\}$ appears or disappears, there must be bulk modes that are in one-to-one correspondence with, and which have the same spectral gaps as, the edge modes represented by $U(\{\alpha_j\})$. From our previous discussion following Eq.~\eqref{eq:general_abelian_ueff}, eigenstates of the bulk spectrum at the phase transition point must be accompanied by other modes at quasienergies which are $n\alpha_j$ apart. In general we may consider phase transitions of the type $\{\alpha_j\}\to\{\beta_j\}$ by decomposing them into sequences of phase transitions of the type considered above.

\subsection{Bulk Topological Invariant}

We now consider evolving the closed system through a loop so that the unitary evolution operator at the end of the evolution is $\mathbb{I}$ up to a phase factor $e^{i\phi}$, which we take to be $1$ without loss of generality below. If we now evolve the same system with $\tbb_j$ twisted boundary conditions, then we claim that the unitary operator at the end of the evolution has to be of the form $ e^{ i \theta_j } \mathbb{I}$. 
 
 To see this, consider the following argument:
 the $\tbb_j$-twisted periodic boundary conditions can be applied by transforming the Hamiltonian term $H_{\mathrm{edge},i}$ across a particular link $i$ of the closed chain as $ X_{L(i),j}H_{\mathrm{edge},i}\left(X_{L(i),j}\right)^\dagger $, where $L(i)$ indicates the left side of an edge cut at site $i$.
 
 The Hamiltonians obtained by performing these transformations at different sites $i$ of the chain are related by unitary transformations. In the cases we consider, the unitary loop can be regarded as a finite depth quantum circuit (or an LU transformation) \cite{Chen:2013fo}. The effect of this transformation in a sufficiently long chain at a particular site $i$ must lead to an effective unitary which is of the form $ U_i  $, where $U_i$ acts nontrivially only in the vicinity of $i$ \cite{Lieb:1972wy}.
 
 The unitary obtained by carrying out this transformation at a distant site $j$, $U_j$, must be related to $U_i$ through a unitary transformation. Thus, $U_i$ must be featureless and can only contribute an overall phase---i.e. it must be of the form $ e^{ i \theta_j }\mathbb{I}$.
 
Further, introducing $n_j$ such twists at distant points along the chain must have the same effect as introducing a $(\tbb_j)^{n_j}$ twist, which brings us back to the untwisted case. Thus, $ e^{i n\theta_j} = 1$, and the discrete allowed values of $\theta_j$ imply that it is a topological invariant that cannot change under continuous deformations of the loop. Thus, the set $\{\theta_j\}$ may be used as a bulk topological invariant for the dynamical phase of the system.  The number of distinct invariants is clearly equal to the number of distinct irreducible representations of the group $G$. It also follows that if we consider two loops with invariants $\{ \theta_j \}$ and $\{\beta_j\}$, the loop obtained by running these unitary evolutions in succession has an invariant $\{\theta_j + \beta_j\}$.

A similar argument can be used to confirm that for a general unitary loop of fermions with class D symmetry, $U_0 U^\dagger_\pi=e^{ik\pi} \mathbb{I}$ for integer $k$, as we found in Sec.~\ref{sec:ClassD}. For those systems, a $\pi$ flux insertion can be effected by the appropriate Peierls substitution for a set of edge terms in the vicinity of some site $i$.  Again, we can argue as we did above that this leads to a unitary that differs from that of the closed loop only in the vicinity of $i$ and that this difference must be an overall phase. To argue that the phase must be an integer multiple of $\pi$, we note that introducing $2\pi$ flux brings us back to the case with no flux insertion and that the phase accompanying the effective unitary must have doubled compared to the $\pi$ flux case.

\subsection{Bulk-edge Connection}
Consider a unitary loop with bulk topological invariant $\{\theta_j\}$, which we evolve as follows. We first evolve the system with open boundary conditions along the entire loop. This leads to some effective unitary $U_{\rm eff}$ at time $T$. We then reconnect the edges with some Hamiltonian $H_{\mathrm{ edge}}$ and evolve it for some additional time $T'$, such that the net unitary at $T+T'$ is the identity. This implies that 
\beq
e^{-iH_{\rm edge}T'}U_{\rm eff}=\mathbb{I}.\label{eq:edge_unitary}
\eeq
We may imagine a sequence of moves through which we continuously deform the original loop of the closed system to the above evolution, thus preserving the bulk invariant. If we now consider the related evolution where we evolve it with open boundary conditions until time $T$ and with $\mathbb{T}_j$-twisted boundary conditions from time $T$ to $T+T'$, the unitary at the end of the evolution must be 
\beq
X_{L,j}e^{-iH_{\rm edge}T'}X_{L,j}^\dagger U_{\rm eff}=e^{i\theta_j}\mathbb{I}.
\eeq
This can be rewritten using Eq.~\ref{eq:edge_unitary}  as
\beq
X_{L,j}^\dagger U_{\rm eff} X_{L,j} =e^{i\theta_j}U_{\rm eff},
\eeq
which has the same form as Eq.~\eqref{eq:general_abelian_ueff}, with $\{\theta_j\}$ identified with $\{\alpha_j\}$. This argument also shows us that any effective edge unitary has to satisfy an equation of the form in Eq.~\eqref{eq:general_abelian_ueff}, as claimed before.
 
We can also use this thought experiment to confirm the bulk-edge connection in the class D case. We imagine evolving the system with open boundary conditions along the entire loop, resulting in an effective unitary, $U_{\rm eff}$ at time $T$. We then reconnect it with some edge Hamiltonian $H_{\mathrm{edge}}$, as in our previous discussion, and evolve it until some further time $T+T'$ such that the final unitary of the entire evolution is the identity, leading to the condition of Eq.~\eqref{eq:edge_unitary}. As before, we argue that we can continuously deform the original evolution of the closed system through the loop to the one just described. The effect of redoing the evolution until $T+T'$, but now with $\pi$ flux, must then lead to the unitary : 
  \beq
P_{L}e^{-iH_{\rm edge}T'}P_{L} U_{\rm eff}=e^{ik \pi}e^{-iH_{\rm edge}T'} U_{\rm eff}.
\eeq
where $P_{L}$ is a product of on-site parity operators on the left edge of the system. This can be rearranged to give
\beq
P_L U_{\rm eff} P_L = e^{ik \pi} U_{\rm eff},
\eeq
 which shows that for a non-trivial loop, $U_{\rm eff}$ must satisfy Eq.~\eqref{eq:classd_eff_unit_char} and must commute with $P_L$ for a trivial loop. 
  
\section{Discussion\label{sec:Discussion}}
In this article, we have considered the classification of Abelian Floquet SPT phases in one dimension. Using systems with class D symmetry as an intuitive example, we went on to discuss bosonic SPT phases with $\zbb_N\times\zbb_N$ symmetry and then, more generally, symmetry under any finite Abelian group. Our approach was to consider unitary evolutions as a composition of a constant Hamiltonian evolution with a loop. The first component, being in one-to-one correspondence with a static Hamiltonian, reproduced the classification scheme of static SPT phases. More interesting dynamical SPT phases could be obtained by constructing nontrivial loop evolutions.

In the cases discussed in this paper, we constructed explicit models that could be used to generate such nontrivial loops, and outlined how these models may be extended to more general symmetry groups. The loop construction allowed us to identify dynamical topological order by studying the behavior at the ends of an open system, which we encoded in an effective unitary that acted only at the edge. Effective edge unitaries of this form are distinguished by a discrete set of characteristic transformation properties that are protected from local perturbations. This leads to a spectral pairing for the many-body eigenstates provided the effective edge unitary commutes with the unitary of the bulk, as one might expect for MBL systems.

For closed systems, we argued that unitary loops could only change by certain quantized phases under group-twisted boundary conditions. We  interpreted these phases as  bulk invariants: robust quantities, stable under local perturbations, that  could be used to count and label the dynamical Floquet SPT phases. The distinct  sets of U(1) phases obtained under such twists correspond to distinct one-dimensional irreducible representations of the group. Moreover, these bulk invariants were shown to be related to the effective edge unitary, providing a bulk-edge correspondence in these systems. We argued that invariants of this form can only change through a bulk phase transition, and we enumerated the distinctive properties and gap structures that arise in the quasienergy spectrum at such points.

Overall, we expect that unitary evolutions protected by symmetry $G$ are classified according to the product $n(L) \times H^2(G,U(1))$, where the second factor classifies the static component, and the first factor classifies the loop component. Distinct loop components are given by the different irreducible representations of the group $G$. It is likely that the loop construction would also permit the classification of dynamical SPT phases under more general conditions than those discussed here. We leave a discussion of more exotic symmetries, including higher dimensions, time-reversal symmetry and non-Abelian groups, to future work.

Finally, we note that our classification method and discussion differs from related work \cite{vonKeyserlingk:2016vq,Else:2016tj,Potter:2016tb} in a number of significant ways (the formulation in terms of loops, the bulk invariants, the bulk-edge correspondence and dynamical phase transitions), although the counting of the phases is consistent with these other results. Notably, the product structure of our classification is reminiscent of the product structure of the classification given in Refs.~\onlinecite{Else:2016tj,Potter:2016tb} for $G$ with unitary symmetries only. We expect these classes of SPT phases to be in one-to-one correspondence.  

Although we frame the discussion in the context of Floquet systems, our approach does not require the system to be time periodic. Rather, we consider the unitary evolution due to a general time-dependent Hamiltonian, and study the instantaneous topological order that may exist at a given point in time. In order to obtain the loop component from a general unitary evolution, we assumed that in the closed system, the end-point unitary could be written as $U(T)=\exp(- i H_FT)$, with $H_F$ a static and local Hamiltonian. While this is likely true for many systems exhibiting MBL, it may also hold more generally. In this way, we did not need to explicitly invoke ideas from MBL, although such considerations would undoubtably be important in periodically driven systems. Notably, experimental probes of SPT order, such as the observation of persistent edge modes \cite{Chandran:2014dk,Bahri:2015ib}, would seem to require MBL. We leave a discussion of these issues to future work.

\begin{acknowledgments}
We thank C.~W.~von Keyserlingk and S.~L.~Sondhi for sharing a copy of Ref.~\cite{vonKeyserlingk:2016vq} in advance of publication and for many fruitful discussions. We also thank P.~Padmanabhan and A.~Vishwanath for sharing useful insights and the Princeton Centre for Theoretical Science for hospitality. R.~R. and F.~H. acknowledge support from the NSF under CAREER DMR-1455368 and the Alfred P. Sloan foundation.
\end{acknowledgments}

\textit{Note} --- During the course of this work we became aware of related works in Refs.~\cite{vonKeyserlingk:2016vq,Else:2016tj,Potter:2016tb}, whose results are consistent with those presented here.

\appendix
\section{Further Details on the Class D Model\label{app:classD_details}}
In this appendix, we give additional details on the calculations outlined in Sec.~\ref{sec:ClassD}. We begin with the Hamiltonian of Eq.~\eqref{eq:classD_Ham}, and consider the evolution described in the main next. In the first part of the evolution, for $0\leq t\leq \pi$, the Hamiltonians for open and closed systems are the same, and the unitary evolution operator is given by
\be
U^{o/c}_1(\pi)&=&\exp\left[-\frac{\pi}{2}\left(\sum_{j=1}^K \left(\gamma^a_{2j-1}\gamma^a_{2j}\right)+2\sum_{j=1}^K \left(\gamma^b_{2j-1}\gamma^b_{2j}\right)\right)\right]\\
&=&\prod_{j=1}^K\left(\gamma^a_{2j-1}\gamma^a_{2j}\right),
\ee
where we have made use of the identity
\be
\exp\left[t\gamma_j\gamma_k\right]&=&\cos\left(t\right)+\sin\left(t\right)\gamma_j\gamma_k.
\ee
The second part of the evolution differs depending on whether we are considering the closed or open system. For the closed system, we find
\be
U_2^{c}(\pi)&=&\exp\left[\frac{\pi}{2}\left(\sum_{j=1}^K \left(\gamma^a_{2j}\gamma^a_{2j+1}\right)+2\sum_{j=1}^K \left(\gamma^b_{2j}\gamma^b_{2j+1}\right)\right)\right]\\
&=&\prod_{j=1}^K\left(-\gamma^a_{2j}\gamma^a_{2j+1}\right).
\ee
while for the open system we find
\be
U_2^{o}(\pi)&=&\prod_{j=1}^{K-1}\left(-\gamma^a_{2j}\gamma^a_{2j+1}\right)\\
&\equiv&\left(\gamma^a_{2K}\gamma^a_1\right)\prod_{j=1}^{K}\left(-\gamma^a_{2j}\gamma^a_{2j+1}\right).
\ee
Then, for the complete closed loop, we obtain
\be
U_2^{c}(\pi)U_1^{c}(\pi)&=&\prod_{j=1}^K\left(-\gamma^a_{2j}\gamma^a_{2j+1}\right)\prod_{j=1}^K\left(\gamma^a_{2j-1}\gamma^a_{2j}\right)\\
&=&-1,
\ee
making use of the identity
\be
\prod_{j=1}^{2K-1} \gamma_j\prod_{j=1}^{2K-1} \gamma_j&=&(-1)^{K+1}.
\ee
For the complete open loop, on the other hand, we obtain
\be
U_2^{o}(\pi)U_1^{o}(\pi)&=&-\gamma^a_{2K}\gamma^a_1\equiv U_{\rm eff}(2\pi)U_2^{c}(\pi)U_1^{c}(\pi).
\ee
It may be verified that $U_{\rm eff}(2\pi)=\gamma^a_{2K}\gamma^a_1$ is equivalent to the expressions given for the effective edge unitary in Eq.~\eqref{eq:classD_eff_unit}.

We now consider threading $\pi$-flux through the closed system as suggested at the end of Sec.~\ref{sec:ClassD_singlep}. In the first part of the drive, this does not affect the unitary evolution, since the Hamiltonian only consists of on-site terms. In the second part of the drive, we implement the new boundary conditions by changing the sign of the term that connects site $K$ to site $1$. Specifically, the Hamiltonian for the second part of the drive becomes
\be
\tilde{H}_{2a}&=&\frac{i}{2}\sum_{j=1}^{K-1}\left(\gamma^a_{2j}\gamma^a_{2j+1}\right)-\frac{i}{2}\gamma^a_{2K}\gamma^a_1,
\ee
and similarly for $\tilde{H}_{2b}$. This now corresponds to the unitary evolution
\be
\tilde{U}_2^{c}(\pi)&=&\exp\left[\frac{\pi}{2}\left(\sum_{j=1}^{K-1} \left(\gamma^a_{2j}\gamma^a_{2j+1}\right)-\gamma^a_{2K}\gamma^a_1\right.\right.\\
&&\left.\left.+2\sum_{j=1}^{K-1} \left(\gamma^b_{2j}\gamma^b_{2j+1}\right)-2\gamma^a_{2K}\gamma^a_1\right)\right]\\
&=&-\prod_{j=1}^K\left(-\gamma^a_{2j}\gamma^a_{2j+1}\right).
\ee
Finally, for the complete closed loop, we obtain
\be
\tilde{U}_2^{c}(\pi)U_1^{c}(\pi)&=&-\prod_{j=1}^K\left(-\gamma^a_{2j}\gamma^a_{2j+1}\right)\prod_{j=1}^K\left(\gamma^a_{2j-1}\gamma^a_{2j}\right)\\
&=&+1,
\ee
which has an overall phase difference of $e^{i\pi}$ compared to the flux-free system.

\section{Integer-valued SPT Hamiltonians for Systems with $\zbb_N\times\zbb_N$ Symmetry\label{app:ZNZN}}
In this appendix, we outline the construction of integer-valued $\zbb_N\times\zbb_N$ SPT Hamiltonians. We start from the model $\mathbb{Z}_N\times\mathbb{Z}_N$ Hamiltonians given in Ref.~\cite{Geraedts:2014ts}. These consist of a chain of $Z_N$ variables, divided into odd and even sites, which have the on-site terms
\beq
H^{i,c}_o&=&-\frac{1}{2}\left\{\left(Z^\dagger_{i-1}\right)^cX_i\left(Z_{i+1}\right)^c+\hc\right\}\label{eq:onsite_term}\\
H^{i,c}_e&=&-\frac{1}{2}\left\{\left(Z_{i-1}\right)^cX_i\left(Z^\dagger_{i+1}\right)^c+\hc\right\},\nonumber
\eeq
where $o$ $(e)$ stands for odd (even) sites. In these expressions, $Z$ and $X$ are $\mathbb{Z}_N$ generalisations of the Pauli sigma matrices, given explicitly in Ref.~\cite{Morimoto:2014bz}. Different values for $c$ (modulo $N$) allow one to obtain the $N$ different static SPT phases, and the generators of the $\mathbb{Z}_N\times\mathbb{Z}_N$ symmetry are given by the operators
\beq
V(\mathbb{T}_o)&=&\prod_{i\in o}X_i\nonumber\\
V(\mathbb{T}_e)&=&\prod_{i\in e}X_i,
\eeq
where $\mathbb{T}_o$ and $\mathbb{T}_e$ are the generators of the global $\mathbb{Z}_N$ symmetries on the odd and even sites, respectively. 
We emphasize that each term in Eq.~\eqref{eq:onsite_term} commutes with any other term with the same value of $c$, and that that these also commute with the symmetry generators $V(\mathbb{T}_o),V(\mathbb{T}_e)$.

Each term $H^{i,c}_{o/e}$ has eigenvalues
\begin{equation}
E=-1,-\left(\frac{\omega+\omega^*}{2}\right),-\left(\frac{\omega^2+\left(\omega^*\right)^2}{2}\right),\ldots
\end{equation}
where $\omega=e^{2\pi i/N}$ is a root of unity. We can therefore rewrite $H^{i,c}_{o}$ as
\begin{equation}
H_o^{i,c}=-\left[P_{0,i}+\ldots+\frac{\omega^n+\left(\omega^*\right)^n}{2}P_{n,i}+\ldots\right],
\end{equation}
and similarly for the even sites, where $P_{n,i}$ is the projector to the eigenstate subspace of $H_o^{i,c}$ with eigenvalue $-\frac{\omega^n+\left(\omega^*\right)^n}{2}$. It is now possible to write down modified on-site terms through
\beq
H_{\mathrm{int},o}^{i,c}&=&\sum_{n=1}^{N-1}n P_{n,i}\nonumber\\
H_{\mathrm{int},e}^{i,c}&=&\sum_{n'=1}^{N-1}n' P_{n',i},
\eeq
which each have integer eigenvalues $E=0,1,\ldots,N-1$.

The complete modified integer-valued Hamiltonians are written
\beq
H_{\rm cl}^c&=&\sum_{i\in o}H_{\mathrm{int},o}^{i,c}+\sum_{i\in e}H_{\mathrm{int},e}^{i,c}\nonumber\\
H_{\rm op}^c&=&\sum_{i\in o}'H_{\mathrm{int},o}^{i,c}+\sum_{i\in e}'H_{\mathrm{int},e}^{i,c},
\eeq
where the primes on the summations in the open system indicate that the boundary terms (corresponding to $i=1$ and $i=K$) should be excluded. Below, we will leave the open/closed label implicit if the meaning is clear.

Since each term in the Hamiltonian commutes, we can label each eigenstate of the system by its eigenvalues under each $H^{i,c}_{\mathrm{int},o/e}$. For the closed system, we write a particular state as $\ket{\lambda_1,\lambda_2,\ldots,\lambda_K}$, where each $\lambda_i$ takes values from $\{0,1,\ldots,N-1\}$. Since there are $N^K$ states and $N^K$ eigenvalue combinations, every state is fully determined by its labels. It is simple to verify that the state $\ket{\{\lambda_i\}}$ has energy $E_{\{\lambda_i\}}=\sum_i\lambda_i$ and is transformed under the symmetry generators according to
\beq
V(\mathbb{T}_o)\ket{\{\lambda_i\}}&=&e^{(2\pi i/N)\sum_{i\in o}\lambda_i}\ket{\{\lambda_i\}}\nonumber\\
V(\mathbb{T}_e)\ket{\{\lambda_i\}}&=&e^{(2\pi i/N)\sum_{i\in e}\lambda_i}\ket{\{\lambda_i\}}.\label{eq:symmetry_eigenstates}
\eeq

In the open system, there are still $N^K$ states but only $N^{K-2}$ eigenvalue combinations, since $\lambda_1$ and $\lambda_K$ are not specified. This leaves an $N^2$-fold degeneracy  in every eigenstate of the system. However, the degenerate states within a multiplet can be labeled by their eigenvalues under the symmetry generators $V(\mathbb{T}_{o/e})$, which again take the possible values in $e^{i2\pi s_{o/e}/N}$, with $s_{o/e}\in\{0,1,\ldots,N-1\}$. We write the states of the open system as $\ket{\lambda_2,\ldots,\lambda_{K-1};s_o,s_e}$.

More generally, we define a larger set of integer-valued Hamiltonians through
\beq
H^c_{a,b}&=&a\sum_{i\in o}H_{\mathrm{int},o}^{i,c}+b\sum_{i\in e}H_{\mathrm{int},o}^{i,c},
\eeq
where $(a,b)$ are integers taking the values $\{1,2,\ldots,N\}$. These may describe either open or closed systems by refining the range of the summations. We obtain the trivial integer-valued Hamiltonian of this form by setting $c=0$ through $H^{\rm triv}_{(a,b)}=H_{(a,b),\rm cl}^0$. Note this is the appropriate trivial Hamiltonian for both open and closed chains, since trivial (on-site) boundary terms exist in both types of system.

The Hamiltonians $H^c_{a,b}$ have useful properties when used to evolve a system in time. Acting on an eigenstate of the closed system, we find $U(t)$ has the action
\begin{equation}
e^{(-i H^c _{a,b}t)}\ket{\{\lambda_i\}}=e^{\left(-iat\sum_{i\in o}\lambda_i-ibt\sum_{i\in e}\lambda_i\right)}\ket{\{\lambda_i\}}.
\end{equation}
After evolving through a time period $t=2\pi/N$, we find the unitary operator has the simple representation
\beq
U\left(\frac{2\pi}{N}\right)&=&\left[V\left(\mathbb{T}_o\right)\right]^{-a} \left[V\left(\mathbb{T}_e\right)\right]^{-b},
\eeq
by comparison with Eq.~\eqref{eq:symmetry_eigenstates}.

From these, we can construct useful unitaries by evolving with a two-part Hamiltonian
\begin{equation}
H=\left\{\begin{array}{ccc}
-H^{c_1}_{a,b}&&\mbox{for $0\leq t\leq2\pi/N$}\\
-H^{c_2}_{\bar{a},\bar{b}}&&\mbox{for $\frac{2\pi}{N}\leq t\leq4\pi/N$}
\end{array}\right.,
\end{equation}
where $\bar{a}=N-a$ is the inverse integer to $a$ in $\zbb_N$. Unitaries of this form are loops, since the closed system satisfies $U(4\pi/N)=\mathbb{I}$, while the open system may host edge modes at quasienergies that are multiples of $2\pi/N$. For the specific case of $c_1=1,c_2=0$, we find that the open system unitary at the end of the loop affects the sites at the edge of the chain through
\begin{equation}
U\left(\frac{4\pi}{N}\right)=\exp\left[-i\frac{2\pi}{N}\left(aH_1+bH_K\right)\right].
\end{equation}

\end{document}